# Team Size and Its Negative Impact on the Disruption Index

Yiling Lin[1], Linzhuo Li[2*] and Lingfei Wu[1*]

[1] School of Computing and Information, The University of Pittsburgh, 135 N Bellefield Ave, Pittsburgh, PA 15213

[2] Department of Sociology, Zhejiang University, 866 Yuhangtang Road, Hangzhou, Zhejiang, 310058, P.R. China

[*]Corresponding author. E-mail: liw105@pitt.edu (L.W)

**Abstract**

As science transitions from the age of lone geniuses to an era of collaborative teams, the question of whether large teams can sustain the creativity of individuals and continue driving innovation has become increasingly important. Our previous research first revealed a negative relationship between team size and the Disruption Index—a network-based metric of innovation—by analyzing 65 million projects across papers, patents, and software over half a century. This work has sparked lively debates within the scientific community about the robustness of the Disruption Index in capturing the impact of team size on innovation. Here, we present additional evidence that the negative link between team size and disruption holds, even when accounting for factors such as reference length, citation impact, and historical time. We further show how a narrow 5-year window for measuring disruption can misrepresent this relationship as positive, underestimating the long-term disruptive potential of small teams. Like "sleeping beauties," small teams need a decade or more to see their transformative contributions to science.

## Main

As science transitions from the age of lone geniuses to an era of collaborative teams, a critical question arises: Can large teams sustain individual creativity and continue driving innovation? Our previous research (Wu, Wang, and Evans 2019) was the first to uncover a negative relationship between team size and innovation performance, measured by the Disruption Index—a network-based metric of innovation—through an analysis of over 65 million papers, patents, and software products spanning 1954 to 2014. We further demonstrated that this negative relationship is remarkably robust across thirteen controlled variables for research articles, summarized below (analyses for patents and software are omitted, as this study focuses on research articles):

Table 1. Robustness Tests for the Negative Relationship Between Team Size and the Disruption Index.

| **Index** | **Controlled Variable** | **Explanation** | **Results in Wu et al. 2019** |
|---|---|---|---|
| 1 | Citation Impacts | High-impact vs. low-impact articles, divided into six percentile groups | Main Figure 3 |
| 2 | Academic Disciplines | Nine disciplines, including Physical Sciences, Biology, Medicine, and others | Main Figure 3; Method: "Fields, subfields, and journals of WOS papers" |
| 3 | Publication Years | Articles spanning 1954–2014 | Main Figure 3 |
| 4 | Authors | Numeric index for disambiguated scholars among 38,000,470 unique authors, controlled using fixed-effect regression models | Main Figure 3; Supplementary Information Table 4 |
| 5 | Article topics | 100-dimensional vectors generated using a doc2vec model (Gensim Python library) | Method: "Modeling topics of WOS papers using Doc2vec" |
| 6 | Time Periods | Five distinct historical decades | Extended Data Figure 3 |
| 7 | Time Window Length | Five distinct disruption measurement time windows | Extended Data Figure 3 |
| 8 | Academic Journals | 15,146 academic journals in the Web of Science | Extended Data Figure 4 |
| 9 | Article Types I | 1,502 theoretical articles vs. 2,756 methodological articles | Extended Data Figure 5 |
| 10 | Article Types II | 22,672 reviewing articles vs. 1,338,808 reviewed articles | Extended Data Figure 5 |
| 11 | Self-citation removal | Citations referencing papers with at least one common author are removed (account for 10.2% of all 615,697,434 citations) | Extended Data Figure 5; Method: "Removing self-citations from WOS papers" |
| 12 | Alternative Metrics | Five alternative definitions of disruption | Extended Data Figure 5 |
| 13 | Awards | 877 Nobel Prize-winning papers compared to a control group | Extended Data Figure 10 |

Recent research reported a positive marginal effect of team size on the D-index while accounting for a set of control variables (Petersen, Arroyave, and Pammolli 2024b, 2024a), contrary to our findings (Wu, Wang, and Evans 2019). This raises questions about whether the negative relationship we observed could be attributed to a specific combination of confounders not tested in our 2019 study. One concern raised in Petersen et al. relates to self-citations, which they suggest might inflate the observed disruption scores. However, our 2019 study already addressed this issue by explicitly removing self-citations, a control summarized in Table 1 (Index 11). Given this, we focus on a more consequential difference between our study and Petersen et al.: the choice of the citation window.

Upon closely examining their model, we find that the key difference lies not in the confounders but in the dependent variable—the D-index. While we used the longest available citation window for the D-index across all papers in our dataset, their reliance on a short 5-year window likely introduces bias. We explain this in detail below.

Small teams act as "sleeping beauties," requiring more time to accumulate citations compared to large teams. Extended Data Fig. 7 in Wu et al. (2019) illustrates the difference in long-term citation dynamics between small and large teams and demonstrates a positive correlation between the Disruption Index and the Sleeping Beauty Index (Ke et al. 2015). Since the D-index stabilizes only after a paper stops receiving citations, using a short citation window overestimates the disruptive impact of large teams while underestimating it for small teams. This leads to a mistaken observation of a positive effect of team size on the Disruption Index. In the text below, we first replicate their results and then demonstrate how the negative relationship between team size and the Disruption Index reappears as the citation window length increases.

The same authors published two models (Petersen, Arroyave, and Pammolli 2024b, 2024a), but their specifications and sample selection criteria appear to vary across studies. We begin by revisiting their first model on the team size effect, denoted as Eq. 3 in (Petersen, Arroyave, and Pammolli 2024b), which we cite and re-index below to assess its consistency and appropriateness for evaluating the relationship between team size and disruptiveness.

$$D_{p,5} = b_0 + b_k lnk_p + b_r lnr_p + b_c lnc_p + D_t + \epsilon_t \quad \text{Eq. (1)}$$

Here, $D_{p,5}$ represents the Disruption Index of paper $p$, calculated using a five-year citation window; $k_p$ is the team size (number of coauthors); $r_p$ is the reference length; and $c_p$ is the number of citations. $D_t$ denotes the yearly fixed effects of the Disruption Index. The independent variables were modeled using a logarithmic transform to address their right-skewed distribution.

To fit this model, we use the same dataset—the archived version of Microsoft Academic Graph (now integrated into OpenAlex)—as Petersen et al., and apply the same selection criteria: $1 \leq k_f \leq 10$ for coauthors, $5 \leq r_f \leq 50$ for references, and $10 \leq c_f \leq 1{,}000$ for citations. While Petersen et al. calculated the 5-year disruption index for all 3 million papers published between 1990 and 2009 that met these criteria, we selected five cohorts of papers, totaling 1.9 million, published in 1995, 2000, 2010, 2015, 2017, and 2019. This selection allows us to calculate the disruption index for these papers up to 2020, representing citation windows of 25, 20, 10, 5, 3, and 1 year. We chose this approach because we hypothesize that citation windows moderate the relationship between team size and disruption index and aim to test this hypothesis.

Like Petersen et al., we performed ordinary least squares (OLS) estimation using the STATA 13.0 package *xtreg*. Using the same model, dataset criteria, and software, we successfully replicated their Fig. 2g (Petersen, Arroyave, and Pammolli 2024b), which appears to show a positive marginal effect of team size on the 5-year D-index (Fig. 1c).

However, this result is influenced by the use of a short citation window. When applying the same regression model to other cohorts, we observe that the negative impact of team size on the Disruption Index clearly reemerges as the citation window increases, with the turning point occurring at a 10-year citation window (Fig. 1a-f).

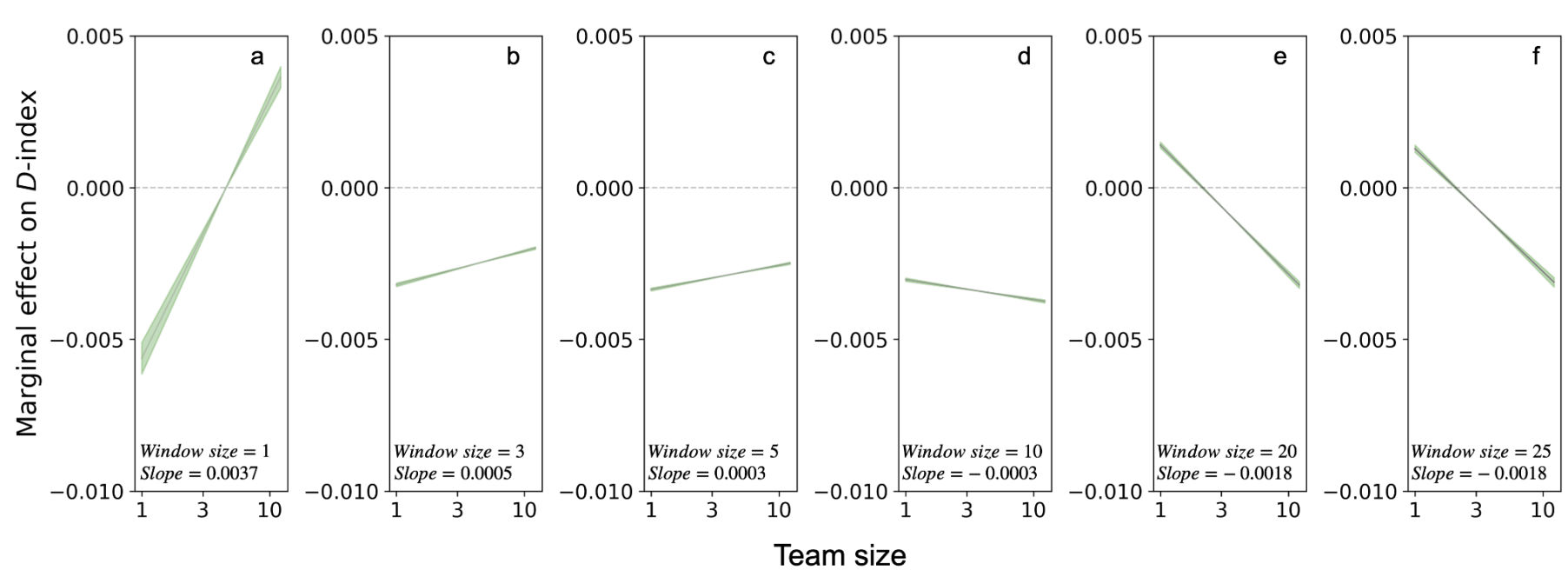

**Figure 1. The negative impact of team size on the D-index reappears with long-term citations.** To examine how citation window length moderates the relationship between team size and the D-index, we analyzed six annual cohorts of papers, each receiving citations from subsequent papers published through 2020. Our dataset includes 47,129 papers from 2019, 271,496 from 2017, 444,675 from 2015, 536,463 from 2010, 344,582 from 2000, and 226,358 from 1995, corresponding to citation windows of 1, 3, 5, 10, 20, and 25 years, respectively. The regression coefficients (slopes) estimated from Eq. 1 are presented, with marginal effects calculated while holding all other covariates at their mean values. Light green confidence intervals are shown around the regression lines.

Next, we re-analyze their second model, which includes additional nonlinear effects and is denoted as Eq. 5 in (Petersen, Arroyave, and Pammolli 2024a), cited and re-indexed below:

$$D_{p,5} = b_t + b_r lnr_p + b_c lnc_p + b_{r2}(lnr_p)^2 + b_{c2}(lnc_p)^2 + \gamma_k + \epsilon_t \qquad \text{Eq. (2)}$$

Building on the variables in Eq. 1—$k_p$ for team size, $r_p$ for reference length, and $c_p$ for the number of citations—Eq. 2 introduces two key changes. First, it adds quadratic terms for $r_p$ and $c_p$ to better capture their nonlinear impacts on the D-index. Second, it represents team size as a set of dummy variables (e.g., whether the team size is 3, 4, or 5), denoted by $\gamma_k$, to more accurately estimate the marginal effect of team size on the D-index for each team size value.

To fit this model, we apply the same selection criteria as in (Petersen, Arroyave, and Pammolli 2024a) to select papers from the Microsoft Academic Graph: $1 \leq k_f \leq 25$ for coauthors, $10 \leq r_f \leq 200$ for references, and $1 \leq c_f \leq 1{,}000$ for citations, and re-run ordinary least squares (OLS) estimation using the STATA 13.0, focusing on two cohorts of papers, 2015 and 2000, corresponding to 5-year and 20-year window size, respectively.

Despite the added complexity of this model, we find that the pattern observed in our previous analysis remains consistent. Specifically, while we successfully replicate the positive effect of team size—biased by the use of a short citation window (Fig. 2a)—the negative impact of team size on the Disruption Index re-emerges as the citation window lengthens (Fig. 2b). To further validate the robustness of our findings, we conduct an additional analysis using the same cohort of 676,831 papers published in 2005, comparing the five-year and twenty-year versions of the Disruption Index. This cohort-based analysis produces results nearly identical to Fig. 2a and Fig. 2b, confirming that the observed pattern holds and reinforcing that the effect of team size on disruptiveness is not an artifact of varying sample compositions.

Based on these analyses across two models (Petersen, Arroyave, and Pammolli 2024b, 2024a), the moderating effect of citation window size is clear. Therefore, we did not replicate nuanced versions of the D-index with a 5-year window, such as the year-journal normalized D-index from Petersen, Arroyave, and Pammolli (2024a), as we do not anticipate significant differences.

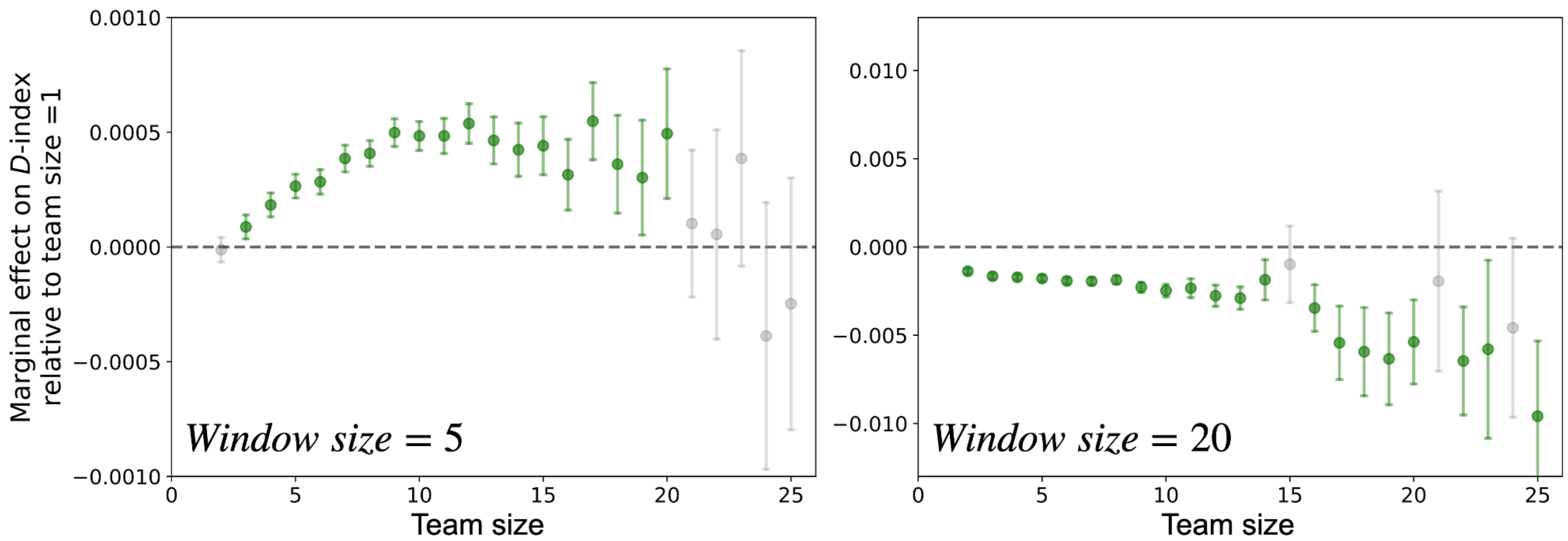

**Figure 2. The negative impact of team size on the D-index reappears with long-term citations, with team sizes modeled as dummy variables.** To examine how citation window length moderates the relationship between team size and the D-index, we analyzed two annual cohorts of papers, each receiving citations from subsequent papers published through 2020. Our dataset includes 1,436,276 papers from 2015 and 459,975 from 2000, corresponding to citation windows of 5 and 20 years, respectively. Dots represent the marginal effects of team size, calculated with all other covariates held at their mean values. Bars indicate 95% confidence intervals, with gray error bars and gray dots denoting results not statistically different from the baseline level ($p > 0.05$), as indicated by the horizontal dashed line.

In sum, the Disruption Index (D-index) evolves over time as the focal paper and its references accumulate citations, stabilizing only after citation growth ceases. Since small teams are like "sleeping beauties," taking longer to gather citations compared to their large counterparts, using a short citation window provides an incomplete picture of the relationship between team size and the D-index, potentially overemphasizing the impact of large teams while underestimating that of small teams (Leahey, Lee, and Funk 2023).

Our findings reaffirm that the negative relationship between team size and the D-index holds when longer citation windows are applied, with the effect stabilizing at windows of ten years or more (Bornmann and Tekles 2019; Lin, Evans, and Wu 2022). This underscores the importance of selecting an appropriate citation window for accurate measurements of innovation. Furthermore, the effect of citation window size is not merely a technical issue. The finding that small teams often require a decade or more to realize their transformative contributions to science underscores the need for science policy and institutional changes to better support small-team science for disruptive innovation. These additional analyses further reinforce the conclusions of our 2019 study on the important role of small teams in driving innovation (Wu, Wang, and Evans 2019).

**Acknowledgments.** We are grateful for support from the National Science Foundation grant SOS: DCI 2239418 (L.W.).